\documentclass[sigconf]{acmart}
\AtBeginDocument{%
  \providecommand\BibTeX{{%
    Bib\TeX}}}

\setcopyright{acmlicensed}
\copyrightyear{2018}
\acmYear{2018}
\acmDOI{XXXXXXX.XXXXXXX}
\acmConference[Conference acronym 'XX]{Make sure to enter the correct
  conference title from your rights confirmation email}{June 03--05,
  2018}{Woodstock, NY}
\acmISBN{978-1-4503-XXXX-X/2018/06}

\usepackage{booktabs}
\usepackage{amsmath} 
\usepackage{algorithmic}
\usepackage{graphicx}
\usepackage{subcaption}
\usepackage{textcomp}
\usepackage{xcolor}
\usepackage[table]{xcolor}
\usepackage{tabularx}
\usepackage{tikz}
\usetikzlibrary{arrows.meta}
\def\BibTeX{{\rm B\kern-.05em{\sc i\kern-.025em b}\kern-.08em
    T\kern-.1667em\lower.7ex\hbox{E}\kern-.125emX}}
\usepackage{multirow} 

\PassOptionsToPackage{hidelinks}{hyperref}
\usepackage{enumitem}

\usepackage{todonotes}
\usepackage{xurl}
\usepackage{listings}
\usepackage{booktabs}

\definecolor{codegray}{rgb}{0.95,0.95,0.95} 
\definecolor{codeblue}{rgb}{0,0,0.6}
\definecolor{codegreen}{rgb}{0,0.5,0}
\definecolor{codered}{rgb}{0.6,0,0}

\begin{document}

\title[Enhanced Qubit Placement and Resource Selection for State-Vector Simulation]{Communication-Aware Qubit Placement and\texorpdfstring{\\}{ } Automatic Node-Count Allocation for\texorpdfstring{\\}{ } Distributed State-Vector Simulation}



\author{Íñigo Aréjula-Aísa}
\orcid{0009-0006-7354-0915}
\affiliation{%
  \institution{Barcelona Supercomputing Center (BSC)}
  \city{Barcelona}
  \country{Spain}}

\author{Sergio Iserte}
\orcid{0000-0003-3654-7924}
\affiliation{%
  \institution{Barcelona Supercomputing Center (BSC)}
  \city{Barcelona}
  \country{Spain}}

\author{Petter Sandås}
\orcid{0009-0009-2028-7719}
\affiliation{%
  \institution{Barcelona Supercomputing Center (BSC)}
  \city{Barcelona}
  \country{Spain}}

\author{Ricard S. Raigada-García}
\orcid{0009-0009-9684-4745}
\affiliation{%
  \institution{Universitat Politècnica de Catalunya (UPC)}
  \city{Barcelona}
  \country{Spain}}

\author{Antonio J. Peña}
\orcid{0000-0002-3575-4617}
\affiliation{%
  \institution{Barcelona Supercomputing Center (BSC)}
  \city{Barcelona}
  \country{Spain}}

\renewcommand{\shortauthors}{Aréjula-Aísa et al.}

\begin{abstract}
Quantum circuit simulation enables the execution of large-scale quantum algorithms by partitioning qubits across multiple computational nodes. However, an inefficient qubit placement leads to excessive inter-node communication, limiting performance and scalability. This work introduces a framework that computes a communication-aware qubit placement offline, minimizing data exchange among nodes, and selects, before circuit execution, the largest node count whose performance degradation stays below a user-defined threshold. The framework is integrated with the QuEST simulator. Evaluations on synthetic benchmarks and real-world QASMBench circuits show that optimized placements reduce communication and achieve runtime speedups of up to 2.3$\times$ over the identity placement, particularly in circuits with sparse, structured interactions. Our results show that qubit placement strongly affects the efficiency of distributed simulation and provide a practical, resource-scalable solution for quantum workloads.
\end{abstract}



\received{20 February 2007}
\received[revised]{12 March 2009}
\received[accepted]{5 June 2009}

\maketitle

\section{Introduction}\label{sec:introduction}
Quantum circuit simulators play a central role in the development and
validation of quantum algorithms on classical hardware, particularly while
large-scale fault-tolerant quantum devices remain unavailable.
State-vector simulators~\cite{jones_quest_2019} are widely used because they
provide a general and exact representation of quantum states. However, their
memory requirements grow exponentially with the number of qubits, making
distributed simulation necessary for circuits beyond the capacity of a single
compute node.

Distributed simulation introduces a trade-off between computational
capacity and communication overhead. When scaling up, simulations could become communication-bound and allocating more nodes does not improve performance, motivating resource-management strategies that select an efficient configuration for each circuit.

This work investigates whether such communication overhead can be reduced in a non-intrusive manner, without modifying the circuit simulator internals. We identify two key factors that govern this overhead: the number of global qubits ($k$), that is, the qubits that determine how the state vector is split across $2^k$ compute nodes, and the assignment of logical qubits to these global positions, referred to as \emph{placement}. Increasing $k$ reduces the memory footprint per node and exposes more parallelism, but could also increase the communications and the volume of exchanged data.

These two aspects give rise to two largely independent but interacting optimization problems. The first is \textbf{placement}, which determines which $k$ logical qubits are assigned to global positions for a fixed node count. While placement does not alter the semantics of the circuit, it directly affects the communication pattern.
The second is \textbf{node count}, which determines how many compute nodes should be used for a given circuit. 
Importantly, the most efficient node count depends not only on the number of qubits but also on the structure of the circuit itself, since different gate patterns induce different communication behaviors. 

The contributions of this work are twofold: (1) a circuit-aware \textbf{placement optimizer} that models circuits as weighted graphs and uses simulated annealing to minimize communication, and (2) a predictive \textbf{node-count efficiency model} that selects, before circuit execution, the largest node count whose predicted performance degradation stays below a user-defined threshold.
We evaluate both approaches using synthetic circuits designed to stress diverse communication patterns, as well as real-world circuits from the QASMBench suite~\cite{li_qasmbench_2023}, on the MareNostrum~5 supercomputer.

\section{Background}\label{sec:background}
This section reviews the three building blocks of this work---distributed state-vector simulation, simulated annealing, and process malleability---and the software used for simulation and malleability in our experiments.

\paragraph{Distributed state-vector simulation.}\label{subsec:background-statevector}
In the state-vector model, an $N$-qubit quantum state is represented
explicitly as a vector of $2^N$ complex amplitudes, each stored as a pair of
floating-point numbers (16 bytes in double precision). For example, a 34-qubit
state requires 256\,GiB, which already exceeds the 256\,GB of memory of a
MareNostrum~5 node.

To overcome this limitation, distributed simulators partition the state vector
across multiple processes. In the configuration considered here, the vector is
distributed across $2^k$ processes, one per compute node, with each process
storing a fraction of the amplitudes. This distribution enables the simulation
of larger circuits and provides parallelism because each process can update
its local portion of the state.

The distribution also introduces communication overhead. The qubits mapped to
the $k$ most significant positions, referred to as \emph{global qubits},
determine how the state vector is partitioned across processes. A gate is
\emph{diagonal} when its unitary matrix is diagonal in the computational basis
(e.g., \texttt{Z}, \texttt{S}, \texttt{T}, \texttt{Rz}, and \texttt{CZ}), so it
only changes the phases of the amplitudes without mixing them. Non-diagonal
gates whose target is a global qubit require data exchanges among processes,
whereas gates that only target local qubits can be applied without
inter-process communication, even when controlled by a global qubit. As more
non-diagonal gates target global qubits, communication
can become a significant component of execution time.

We use QuEST~\cite{jones_quest_2019} as the concrete implementation of this
model in our experiments: it partitions the state vector and applies the gates
as described above.

\paragraph{Simulated annealing.}\label{subsec:background-sa}
Simulated annealing~\cite{kirkpatrick_optimization_1983} is a general-purpose metaheuristic for combinatorial optimization problems. It begins from an initial solution and iteratively explores neighboring candidates, accepting all improvements and, with a probability that decreases over time, some worsening moves. This controlled acceptance of non-improving moves allows the search to escape local minima and improves the chances of finding a low-cost solution.

\paragraph{Process malleability.}\label{subsec:background-malleability}
Process malleability is the capacity of a parallel application to change its number of processes at runtime, acquiring or releasing resources without restarting the job~\cite{dmr_ubs}.
In particular, we use the DMR framework~\cite{iserte_dmrlib_2020}
, which expands or shrinks a running job in coordination with the resource management system. This allows the node count selected by our model to be applied to an already submitted job, before the state vector is created, without resubmitting it.

\section{Related Work}\label{sec:related}
This section reviews prior work related to the two contributions of this paper:
circuit-aware qubit placement and automatic node-count selection.

\paragraph{Qubit placement.}
Several studies reduce communication through circuit-specific placement
decisions. Adamski \emph{et al.}~\cite{adamski_energy_2023} manually
cache-block a QFT circuit in QuEST, moving SWAP operations so that subsequent
Hadamard gates act on local rather than global qubits. This reduces the
fraction of runtime spent in MPI communication from 43\% to 25\%, at the cost
of additional distributed SWAPs. Similarly, De Raedt
\emph{et al.}~\cite{deraedt_universal_2026} report that a manually selected
qubit relabelling in JUQCS-50 reduces the execution time of a 25-bit adder by
approximately 40\%. These results demonstrate the potential of circuit-aware
placement, but the placement decisions are specific to the evaluated circuits
rather than produced by a general automatic search.

More general approaches modify qubit placement dynamically during execution.
Guerreschi~\cite{guerreschi_fast_2022} introduces a compiler pass for Intel-QS
that greedily searches for qubit permutations that reduce the number of
communicating gates and inserts \texttt{PermuteQubits} instructions when
beneficial, reporting approximately a two-fold speedup on random circuits and
larger gains on QFT-like circuits. Teranishi
\emph{et al.}~\cite{teranishi_lazy_2024} (Lazy QR) reduce reordering overhead
in VQE workflows by delaying and merging reorderings through a
time-space-tiling schedule.

A related line of work reduces communication by dynamically partitioning the
circuit. Xu \emph{et al.}~\cite{xu_minimizing_2026} (QuanTrans) group
consecutive layers so that communication is performed once per block rather
than once per layer. Jiao \emph{et al.}~\cite{jiao_communication_2023} combine
gate fusion with on-demand amplitude transfers that are triggered only when the
local qubit assignment changes. In follow-up
work~\cite{jiao_communicationpartition_2026}, they greedily build a complete
partition from each of several candidate starting points, estimate the
communication of each partition with a cost model, and keep the cheapest one.

In summary, circuit-specific studies rely on hand-tuned placements, whereas
dynamic reordering and partitioning approaches make placement decisions, often
greedily, at runtime, and therefore require dedicated support in the simulator
or its execution schedule. In contrast, our placement method formulates the
problem as a weighted graph partitioning task and uses simulated annealing to
minimize an exact cut corresponding to the number of communicating gates. The
result is a single offline placement that applies to arbitrary circuits
and requires neither runtime reconfiguration nor changes to the simulator. Our
method is thus complementary to dynamic approaches, since its placement could
serve as the initial layout from which they start.

\paragraph{Node-count selection.}
Benchmarking studies characterize how distributed simulation scales with the
available resources. Zhong \emph{et al.}~\cite{zhong_scalable_2025} benchmark
scalable CPU/GPU simulation, whereas Brown
\emph{et al.}~\cite{brown_multigpu_2026} analyze the influence of network
performance on multi-GPU simulation. These studies quantify scaling and
interconnect effects, but they do not predict how many nodes a given circuit
should use.

Placement methods, in turn, generally assume that the node count is fixed
externally. Existing studies either tune the number of resources empirically
or scale qubits and GPUs together manually, as in the closeness-centrality
partitioner of Popovici \emph{et al.}~\cite{popovici_closeness_2025}. Tabuchi
\emph{et al.}~\cite{imamura_mpiqulacs_2022} (mpiQulacs) are a partial
exception, but the ratio they use to assess hardware utilization is a post hoc
measure rather than a criterion for selecting the node count before execution.
Our second contribution addresses this gap with an efficiency model that
selects the node count before circuit execution according to a user-defined
performance-degradation threshold.

\section{Methodology}\label{sec:methodology}
This section describes our optimization methods for process communication (Section~\ref{subsec:placement}) and node-count adjustment (Section~\ref{subsec:resources}) in distributed state-vector quantum circuit simulations.

\subsection{Qubit Placement}\label{subsec:placement}
A \emph{layout} is a permutation that assigns each logical qubit to a physical
position in the distributed state-vector representation. With $2^k$ processes,
the $k$ highest-order positions are global, while the remaining positions are
local. Thus, for a fixed $k$, a layout induces a \emph{placement}, that is, the
global set $G$ of $k$ logical qubits; all remaining qubits are local. Since
communication depends only on which qubits are global, layouts that induce the
same placement have the same communication cost, and optimizing the layout
amounts to choosing the placement.

We represent the circuit by a weighted graph whose nodes correspond to logical
qubits. The node weight $w(q)$ counts single-target non-diagonal gates acting
on $q$, while the edge weight $w(a,b)$ counts two-target non-diagonal gates
acting on $a$ and $b$.
These weights directly encode communication: a node contributes when its qubit is global, and an edge contributes once when at least one of its endpoints is global.

Diagonal gates and control qubits are omitted. The former do not require
communication, while, for a controlled non-diagonal gate, communication is
determined by the target qubit rather than by the control. Thus, unlike a
conventional interaction graph, this graph represents gate-target
communication rather than all two-qubit interactions.

Let $\delta^+(G)$ denote the set of weighted edges with at least one endpoint
in the global set $G$. We define the communication cut of a given placement as

\begin{equation}
cut(G)
 =
 \sum_{q \in G} w(q)
 +
 \sum_{e \in \delta^+(G)} w(e).
\label{eq:communication-cut}
\end{equation}

Equivalently, $cut(G)$ counts the number of non-diagonal gates with at least one
target in the global set. The minimum communication cut for a configuration
with $k$ global qubits is therefore

\begin{equation}
cut_{\min}(k)
 =
 \min_{\substack{G \subseteq Q\\ |G|=k}} cut(G),
\label{eq:min-communication-cut}
\end{equation}

where $Q$ denotes the set of logical qubits. 

\begin{figure}[tbp]
\centering
\begin{subfigure}[]{\linewidth}
\centering
\includegraphics[clip,width=0.5\linewidth,trim={0.2cm 0.7cm 0.2cm 2cm}]{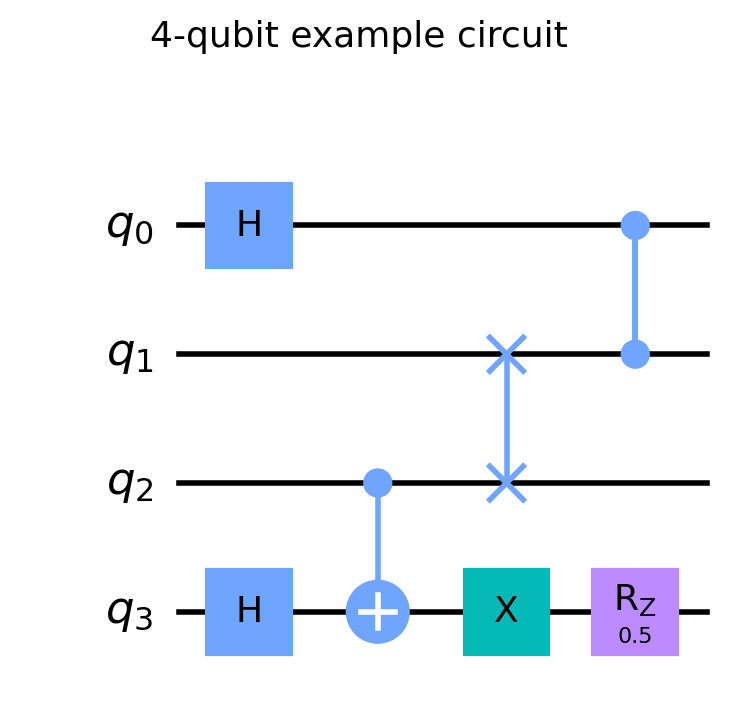}
\caption{Example circuit.}
\label{subfig:example-circuit}
\end{subfigure}
\vspace{0.4em}

\begin{subfigure}[]{\linewidth}
\centering
\includegraphics[clip,width=0.55\linewidth,trim={0.2cm 1.5cm 0.2cm 1cm}]{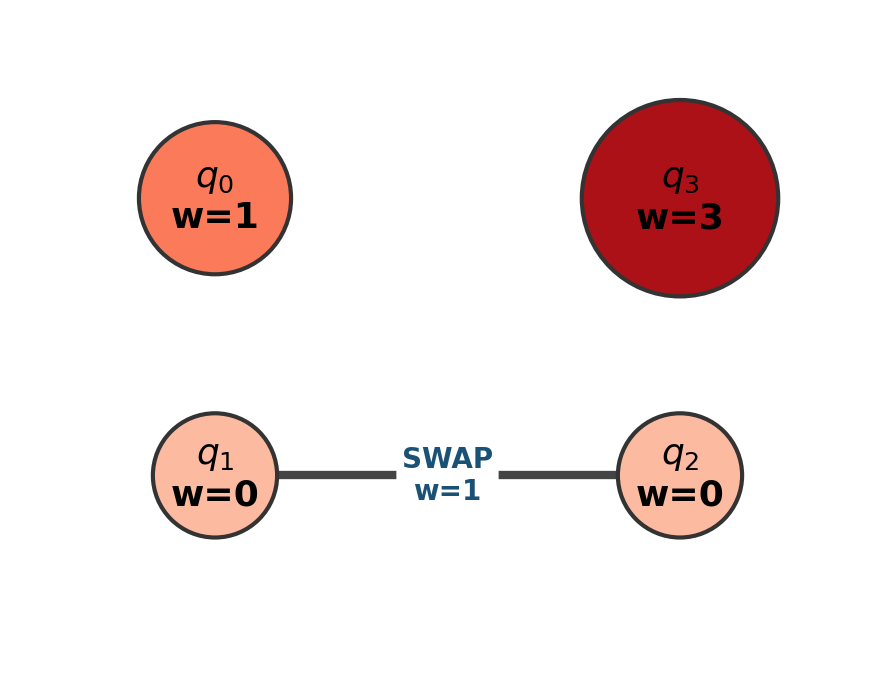}
\caption{Its graph.}
\label{subfig:example-graph}
\end{subfigure}

\vspace{0.4em}

\begin{subfigure}[]{\linewidth}
\centering
\includegraphics[clip,width=0.75\linewidth,trim={0.2cm 1cm 0.2cm 0.75cm}]{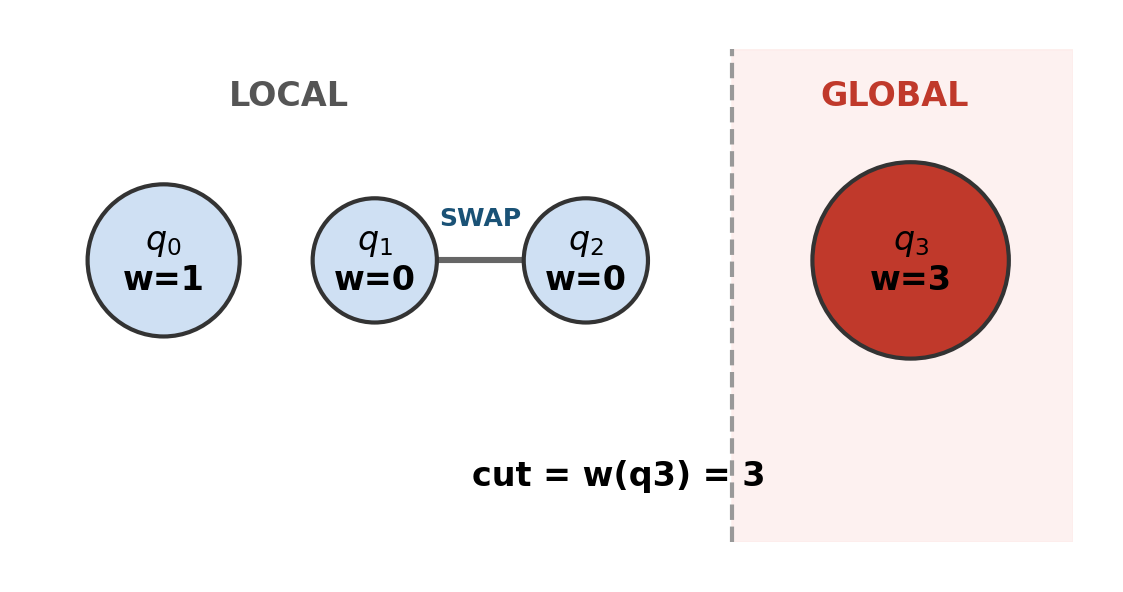}
\caption{Identity layout, cut $=3$.}
\label{subfig:example-cut-identity}
\end{subfigure}
\vspace{0.4em}

\begin{subfigure}[]{\linewidth}
\centering
\includegraphics[clip,width=0.75\linewidth,trim={0.2cm 1cm 0.2cm 0.75cm}]{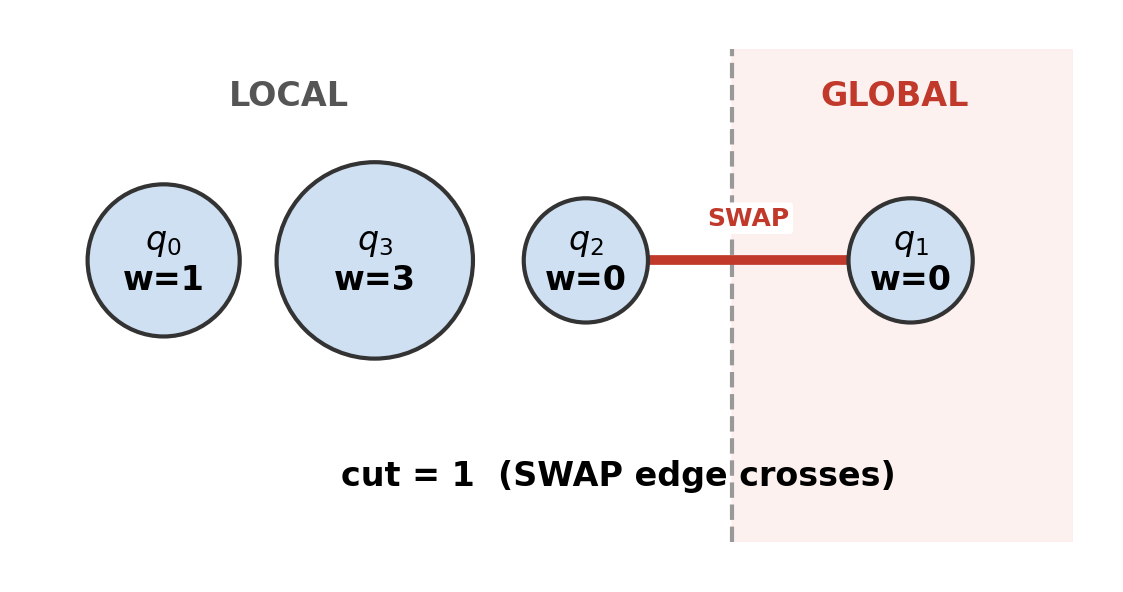}
\caption{Optimized layout, cut $=1$.}
\label{subfig:example-cut-optimised}
\end{subfigure}

\vspace{-0.25cm}
\caption{A 4-qubit worked example at $k=1$. In (c) and (d), red color marks what communicates under that layout: a red node is a global qubit whose non-diagonal gates must communicate, and a red edge is a two-target gate that straddles the local/global boundary; the cut is the sum of both.}
\label{fig:cut-example}
\Description{Four stacked panels. (a) A four-qubit circuit with Hadamard, CNOT, X, Rz, SWAP and CZ gates on qubits q0 to q3. (b) The interaction graph of the circuit, with one node per qubit and one edge per two-qubit gate. (c) The identity layout with one global qubit, where the highlighted nodes and edges give a communication cut of 3. (d) An optimized layout with the same circuit, where the cut drops to 1.}
\vspace{-6pt}
\end{figure}

Figure~\ref{fig:cut-example} illustrates the communication-cut construction using a four-qubit circuit with $k=1$.
The circuit in Figure~\ref{subfig:example-circuit} applies Hadamard gates \texttt{H} to $q_0$ and $q_3$, a controlled-NOT gate \texttt{CNOT} with control $q_2$ and target $q_3$, a Pauli-$X$ gate \texttt{X} on $q_3$, a rotation \texttt{Rz(0.5)} on $q_3$, a \texttt{SWAP} gate between $q_1$ and $q_2$, and a controlled-$Z$ gate \texttt{CZ} between $q_0$ and $q_1$.

The graph contains no mark for the \texttt{CZ} gate because it is diagonal in the computational basis. Likewise, the control $q_2$ of the \texttt{CNOT} does not contribute to the graph: only the target of a controlled non-diagonal gate determines whether communication is required.

Under the identity layout (Figure~\ref{subfig:example-cut-identity}), the heaviest qubit $q_3$, targeted by three non-diagonal gates (its \texttt{H}, the \texttt{CNOT} targeting it, and its \texttt{X}), is assigned to the global region, so all three communicate and $w(q_3)=3$.

Figure~\ref{subfig:example-cut-optimised} shows a placement with minimal communication cut, reassigning $q_3$ to the local region and placing the lighter qubit $q_1$ in the global region instead.
Placing $q_0$ global instead would have reached the same cut, and ties like this can be resolved by keeping whichever option the search reaches first.

In this layout, only the \texttt{SWAP} edge crosses the boundary, reducing the cut from three to one, even though that edge now costs one crossing it did not before.
This example shows how placement alone can reduce communication without modifying the circuit or the simulator.

When all non-diagonal gates have a single target, the graph contains no edges and the communication cut is the sum of the node weights of the global qubits. The optimal layout can then be obtained by assigning the $k$ lowest-weight qubits to global positions.

Two-target non-diagonal gates introduce edges whose costs depend on the placement of both endpoints. In this case, independently sorting qubits by node weight is insufficient, and we search for an efficient layout using simulated annealing (Section~\ref{sec:background}). Starting from the identity layout, the search exchanges one global qubit with one local qubit. Moves within either set are omitted because they cannot change $\operatorname{cut}$. The objective is updated incrementally using only the gates affected by the exchange.

The annealing schedule cools geometrically from an initial temperature selected to accept mildly worsening moves with high probability. The search is bounded by a fixed number of iterations proportional to the number of qubits ($N$); since each iteration inspects only the gates touching the two exchanged qubits, the search runs in $O(N \cdot M)$ time, where $M$ is the circuit's number of non-diagonal gates. It retains the best layout found, and a fixed random seed makes the search reproducible. As simulated annealing is a heuristic, the resulting placement is not guaranteed to be globally optimal.

Before execution, the selected layout is applied as a relabeling of qubit indices. This operation moves no amplitudes and leaves the state vector unchanged, apart from the corresponding permutation of its amplitude-index bits. Thus, it preserves circuit semantics while reducing communication.

\subsection{Node-Count Efficiency Model}\label{subsec:resources}

Beyond qubit placement, the second optimization is how large to set $k$, that is, how many of the $2^k$ processes (computational nodes) to use for a given circuit. 
Each node stores a block of $2^{N-k}$ amplitudes, where $N$ is the total number of qubits and $k$ is the number of global qubits, with each amplitude represented by two double-precision floats (16\,bytes). The amplitudes must fit within the per-node memory budget, which imposes a lower bound on $k$:
\[
k_{\text{mem}} = \left\lceil \log_2 \left( \frac{2^N \cdot 16}{mem_{\text{node}}} \right) \right\rceil,
\]
where $mem_{\text{node}}$ is the memory per node. Together with the qubit-count ceiling $k < N$ (a register needs at least one local qubit to distribute meaningfully and can therefore spread over at most $2^{N-1}$ nodes), this gives the feasible range of global qubits
\[
k \in [k_{\text{mem}},\, N-1].
\]

Doubling the node count adds one global qubit, which halves the local block size.
Involving more nodes into the computation can speed up the circuit execution, even at the expense of higher communication overheads.

We formulate this trade-off through a predictive efficiency model. 
For this purpose, we rely on the Communication Efficiency (CE) metric\footnote{\url{https://pop-coe.eu/node/69}},
\[
CE(k) = \frac{T_{\text{comp}}(k)}{T(k)}
       = \frac{1}{1 + O(k)},
\]
where \(T_{\text{comp}}(k)\) is the useful computation time with \texttt{k} global qubits, \(T(k)\) is the total runtime which includes the communication overhead (\(T_{\text{comp}}(k) + T_{\text{comm}}(k)\)), and

\[
O(k) = \frac{T_{\text{comm}}(k)}{T_{\text{comp}}(k)}
\]
is the deviation from linear scaling, in other words, its \textbf{degradation}.
A value $O(k)=0$ (i.e.\ $CE(k)=1$ or $T_{\text{comm}}=0$) means the execution time halves when doubling the nodes.
When $O(k)$ grows above $0$, the communication eats into that ideal doubling. 
Since $CE(k)$ depends only on $O(k)$, we work with $O(k)$ for the rest of the model.

We now make this trade-off analytical.
The compute-only time is $T_{\text{comp}}(k) = CO\cdot GA\cdot 2^{N-k}$, where $GA$ is the number of gates within the circuit and $CO$ is a constant compute time per amplitude per gate.
The communication time $T_{\text{comm}}(k)$ is composed of two terms: a \textbf{latency} term, paid once per communicating gate, and a \textbf{bandwidth} term, proportional to the data volume moved by those gates.
Under our placement model (Section~\ref{subsec:placement}), there are $cut_{\min}(k)$ communicating gates, and their data volume is captured by a communication weight $W_{\text{comm}}(k)$ that accounts for how controls reduce exchanged blocks. This yields
\[
T_{\text{comm}}(k) = \alpha(k)\,cut_{\min}(k) \;+\; \beta(k)\,2^{N-k}\,W_{\text{comm}}(k),
\]
where $\alpha(k)$ and $\beta(k)$ are per-message and per-amplitude costs, respectively. Dividing by $T_{\text{comp}}(k)$ cancels the factor $2^{N-k}$ in the bandwidth term and leaves two per-cluster ratios, $a(k)=\alpha(k)/CO$, in units of amplitudes, and $b(k)=\beta(k)/CO$, dimensionless:
\[
O(k) = a(k)\cdot\frac{cut_{\min}(k)}{GA}\cdot 2^{-(N-k)} \;+\; b(k)\cdot\frac{W_{\text{comm}}(k)}{GA},
\]
where $a(k)$ and $b(k)$ express the cluster's communication cost in units of its own compute speed: $a(k)$ is the number of amplitudes' worth of compute that one message's fixed latency costs, and $b(k)$ is how many times more expensive moving one amplitude is than computing on it. Both are properties of the cluster alone, measured once and reused for every circuit (Section~\ref{subsec:eval-efficiency-model}); $cut_{\min}(k)$, $W_{\text{comm}}(k)$, and $GA$ come from the circuit and its optimized layout alone. This formula lets us predict the efficiency of any circuit without running it, using only a cluster-wide calibration and static analysis of the circuit.

We anchor degradation at the smallest node count the circuit can actually reach, $k_{\min}=k_{\text{mem}}$, rather than at $k=0$, which is not a valid configuration for large $N$:
\[
O_{k_{\min}}(k) = \frac{O(k) - O(k_{\min})}{1 + O(k_{\min})}.
\]
This allows the model to pick, for any circuit, the largest node count whose degradation from its smallest node count stays within a selected threshold $\tau$: the \textbf{efficient} node count,
\[
k_{\text{eff}} = \max\{k \in [k_{\text{mem}}, N-1] : O_{k_{\min}}(k) \leq \tau\}.
\]

\section{Implementation}\label{sec:implementation}
This section describes QuCSLO, the library we developed to implement the placement and node-count methods introduced in Section~\ref{sec:methodology}.

QuCSLO applies the methodology without modifying or reimplementing the simulator itself.
It is organized into five modules:
\begin{itemize}
  \item Circuit Representation (Section~\ref{subsec:impl-circuit}): an ordered sequence of gates stored in a \texttt{Circuit} type.
  \item Circuit Import (Section~\ref{subsec:impl-qasm}): an OpenQASM~2.0 reader that parses circuits into the \texttt{Circuit} type.
  \item Placement Optimizer (Section~\ref{subsec:impl-layout}): the implementation of qubit-layout search described in Section~\ref{subsec:placement}.
  \item Node-Count Selector (Section~\ref{subsec:impl-resources}): the implementation of computational resources selection model described in Section~\ref{subsec:resources}.
  \item Simulator Connector (Section~\ref{subsec:impl-executor}): the integration with the target quantum circuit simulator which ingests the \texttt{Circuit} instance.
\end{itemize}

\begin{figure*}[t]
\centering
\includegraphics[width=0.8\textwidth]{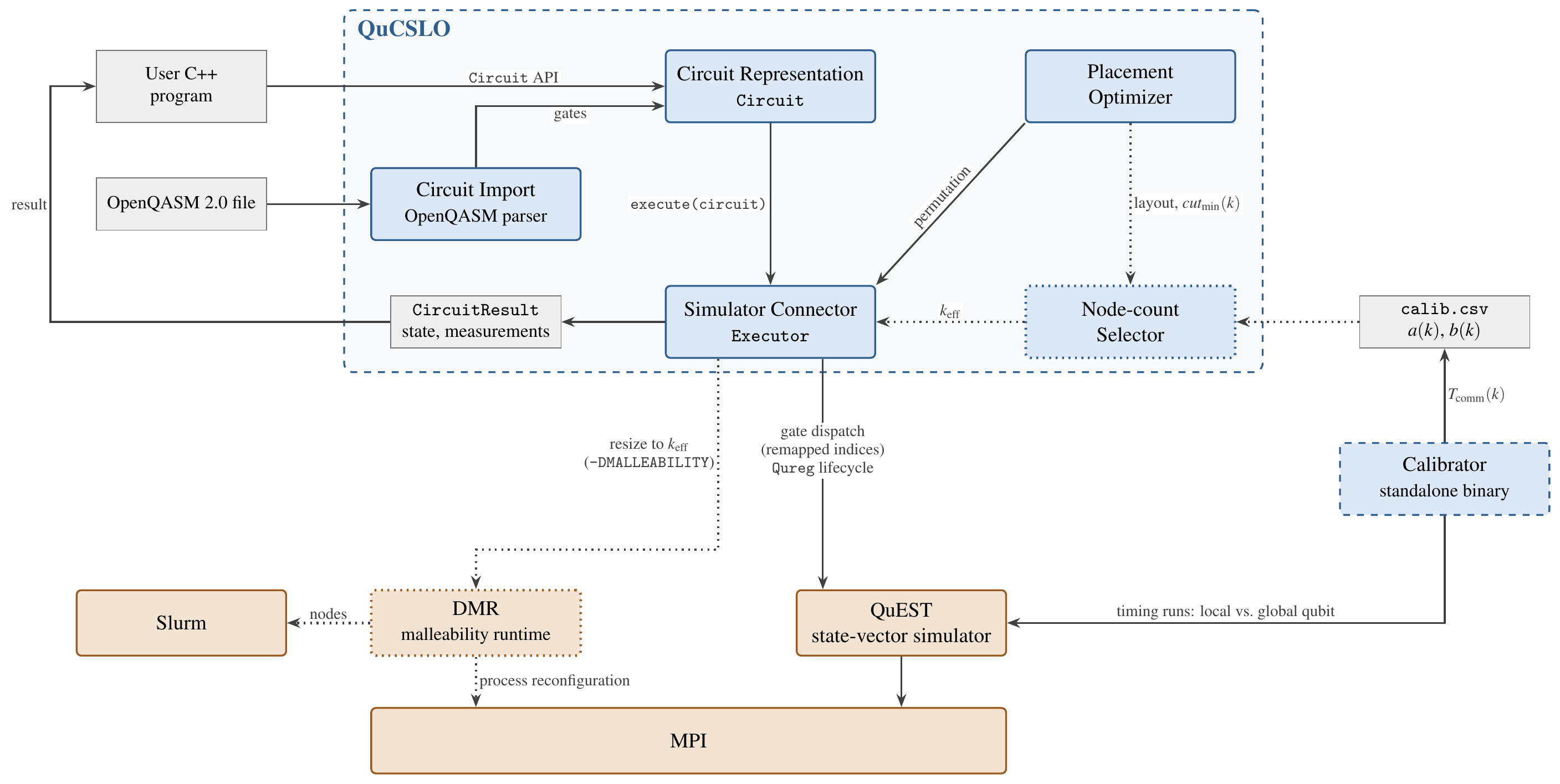}
\vspace{-0.25cm}
\caption{QuCSLO components and their interaction with QuEST, DMR and the system software.}
\label{fig:components}
\Description{Block diagram of QuCSLO. A user C++ program builds a Circuit through the Circuit API, or an OpenQASM 2.0 file is parsed by the Circuit Import module into the Circuit. The Placement Optimizer computes a qubit permutation and the minimum cut per global-qubit count; the Node-Count Selector combines this with calibration coefficients produced by a standalone Calibrator to choose the node count. The Executor dispatches gates with remapped indices to the QuEST state-vector simulator, optionally asks the DMR malleability runtime to resize to the selected node count, and returns a CircuitResult to the user. QuEST and DMR run on top of MPI, and DMR obtains nodes from Slurm.}
\vspace{-12pt}
\end{figure*}

Figure~\ref{fig:components} shows how these modules interact with each other and with the external software.
Blue boxes are QuCSLO components, orange boxes are external software, and grey boxes are the data exchanged between them.
Only the Simulator Connector communicates with QuEST and DMR.
The Calibrator is a standalone binary which runs once per cluster on top of QuEST and writes the \texttt{calib.csv} file read by the Node-Count Selector.
The dashed arrow towards DMR marks the resize path, which is only present when QuCSLO is built with malleability support.

\subsection{Circuit Representation}\label{subsec:impl-circuit}
QuCSLO separates a circuit's description from its execution. 
The \texttt{Circuit} type only records an ordered sequence of gates as data: constructing one does not apply any gate, so the full circuit is available for analysis before any state vector exists. 
It carries no notion of node count or qubit layout either, both of which are properties of how a circuit is run, not what it is.

Qubits are identified by indices, the keys that both the placement optimizer and the communication model index into.

\subsection{Circuit Import}\label{subsec:impl-qasm}
A \texttt{Circuit} can be constructed in two ways: with the \texttt{Circuit} API or with the QuCSLO OpenQASM parser.

On the one hand, the \texttt{Circuit} API provides a programmatic interface for assembling circuits directly in C++. 
Listing~\ref{lst:circuit-api} shows a minimal example that builds and runs a Bell pair. 

\begin{lstlisting}[language=C++, caption={Building and running with the \texttt{Circuit} API.}, label={lst:circuit-api}]
qucslo::Circuit circuit(2);
circuit.H(0);
circuit.CNOT(0, 1);
qucslo::ExecutorConfig config;
config.layout.enabled = true;
qucslo::Executor executor(argc, argv, config);
qucslo::CircuitResult result = executor.execute(circuit);
\end{lstlisting}

The \texttt{Circuit} describes a 2-qubit Bell-state preparation (line~1): a Hadamard on qubit~0 followed by a \texttt{CNOT} from qubit~0 to qubit~1 (lines~2, and~3, respectively). 
The call to \texttt{executor.execute} does not modify the \texttt{Circuit} (line~7); it uses it as a read-only description and returns a \texttt{CircuitResult} that owns the final state and any measurements.

On the other hand, circuits can be imported from OpenQASM~2.0 using the parser provided in QuCSLO.
Listing~\ref{lst:qasm-api} illustrates how \texttt{parseQasmFile} loads a circuit from a file into the same \texttt{Circuit} type.

\begin{lstlisting}[language=C++, caption={Loading and running a circuit from an OpenQASM~2.0 file.}, label={lst:qasm-api}]
qucslo::Circuit circuit = qucslo::parseQasmFile(path);
qucslo::Executor executor(argc, argv);
qucslo::CircuitResult result = executor.execute(circuit);
\end{lstlisting}

The \texttt{parseQasmFile} routine reads an OpenQASM~2.0 file (line~1), resolves its register and gate declarations, and translates each supported operation into the same internal gate representation used by the programmatic API.
Unsupported or non-unitary constructs are rejected, so every downstream pass---placement optimizer and node-count selector alike---sees a uniform gate-level sequence regardless of how the circuit was built.
The resulting \texttt{Circuit} is then passed to the same \texttt{Executor} interface as programmatically constructed circuits (line~3), ensuring that placement, node-count selection, and execution behave identically whether the circuit originates from the API or from an external OpenQASM description. 

\subsection{Placement Optimizer}\label{subsec:impl-layout}
The Placement Optimizer module implements the qubit-layout search. 
It takes a \texttt{Circuit} and a selected number of global qubits \(k\) as input and returns the layout with the lowest communication cut found.

Internally, the module first builds the weighted qubit graph by scanning the circuit once and accumulating node and edge weights from non-diagonal gates, then runs the simulated-annealing search starting from the identity layout (see Section~\ref{subsec:placement}).
This layout is the permutation that maps every logical qubit to its physical position in the state vector.

Line~5 of Listing~\ref{lst:circuit-api} showcases how to enable the optimized layout using the \texttt{Circuit} API and how to apply it to the executor (line~6) before running the simulation (line~7).

\subsection{Node-Count Selector}\label{subsec:impl-resources}
The Node-Count Selector module implements the resource efficiency model. 
Its purpose is to calculate a node count that meets a user-defined quality of service target, expressed as a maximum performance degradation $\tau$, for a given circuit and hardware platform.

Given a \texttt{Circuit} and a set of calibrated cluster parameters, the module first computes the gate count \(GA\), the cut values \(cut_{\min}(k)\), and the communication weights \(W_{\text{comm}}(k)\) over the feasible range of \(k\). 
These circuit-level quantities are then combined with pre-measured latency and bandwidth ratios \(a(k)\) and \(b(k)\) to evaluate the deviation from ideal scaling \(O(k)\) and its anchored variant \(O_{k_{\min}}(k)\) (Section~\ref{subsec:resources}). 
For a user-specified degradation threshold \(\tau\), the module selects the efficient node count \(k_{\text{eff}}\) and exposes it through the \texttt{recommendNodeCount} interface, which for each evaluated \(k\) triggers the placement optimizer. 

The calibration parameters are specific to the cluster on which it is measured; in this regard, QuCSLO provides a tool to calibrate the metrics necessary for the node-count selector.
In particular, the calibrator measures the execution time of a gate in two settings: single process (local qubit) and distributed (global qubit).
The difference among these execution times corresponds to the overhead ($T_{\text{comm}}(k)$) of the distributed scenario, which involves internode network communication.
Although the calibrator evaluates a Hadamard gate, the selected gate is not relevant, since both scenarios are executing the same gate, and that computation time is neutralized.
For each node count \(2^k\), the tool calculates ($T_{\text{comm}}(k)$), \(a(k)\), and \(b(k)\) for the different qubit counts.

Listing~\ref{lst:executor-resize} contains how the output of QuCSLO calibrator is loaded (line~2) to configure the execution of the circuit (line~5).
Furthermore, in order to make effective the process resize, line~3 enables the execution reconfiguration, so resources can be adjusted to the efficiency target.

\begin{lstlisting}[language=C++, caption={Resizing to the model's efficient node count before running.}, label={lst:executor-resize}]
qucslo::ExecutorConfig config;
config.resize.calib = qucslo::loadCommCalib("calib.csv");
config.resize.enabled = true;
qucslo::Executor executor(argc, argv, config);
qucslo::CircuitResult result = executor.execute(circuit);
\end{lstlisting}

\subsection{Simulator Connector}\label{subsec:impl-executor}
The Simulator Connector module integrates QuCSLO with a target state-vector simulator and drives the execution of optimized circuits. 
It isolates simulator-specific details behind a uniform interface.

The central component of this module is the \texttt{Executor} type, which handles the execution lifecycle.
Given a \texttt{Circuit}, a layout permutation, and a selected node count \(k\), the \texttt{Executor} applies the layout, configures the simulator with the requested number of processes, and invokes the simulator's API to run the circuit. 

When QuCSLO is built with malleability support, this module can also interface with DMR to adjust the number of processes and computational nodes at runtime.

In particular, in this paper, the library uses QuEST~\cite{jones_quest_2019} as the simulator: the \texttt{Executor} owns its lifecycle (\texttt{initQuESTEnv} / \texttt{finalizeQuESTEnv} and \texttt{createQureg} / \texttt{destroyQureg}), and dispatch translates each gate in the \texttt{Circuit} into the matching QuEST call (e.g.\ \texttt{applyHadamard}, \texttt{applyControlledPauliX}), relabeling its qubit indices through the layout permutation first. This relabeling is a pure index remap with no data movement, so the simulator itself does not need to be modified.

For the resize enabled by Listing~\ref{lst:executor-resize}, the \texttt{Executor} first consults the node-count model, which also picks the layout for that target node count, and checks with DMR~\cite{sandas2026malleable} whether the current allocation already matches it. If not, it asks DMR to reconfigure the job to the target node count; this replaces the running processes outright, and the new processes re-enter \texttt{main}, reconstruct their own \texttt{Executor}, and create the \texttt{Qureg} distributed over the node count actually reached.

\section{Evaluation}\label{sec:evaluation}
QuCSLO and the implemented methods for placement optimization and node-count selection are evaluated on the MareNostrum~5 (MN5) supercomputer, a pre-exascale system integrated into the EuroHPC-JU infrastructure\footnote{\url{https://bsc.es/marenostrum/marenostrum-5}}.
All experiments are conducted on the General-Purpose Partition (GPP) of MN5. 
Specifically, we utilize a set of 64 nodes equipped with dual Intel Xeon Platinum 8480+ Sapphire Rapids processors (56 cores at 2~GHz per socket, 112 cores per node) and 256~GB of DDR5 memory. The nodes are interconnected via a 100~Gbit/s ConnectX-7 NDR200 InfiniBand network.
\textit{Slurm4DMR} is deployed within an MN5 allocation of resources providing an exclusive environment to manage resources guaranteeing immediate resource availability for reconfigurations~\cite{dmr_ppam24}. 
The proposed methods are compatible with the GPU-accelerated version of QuEST and are ready to be deployed on GPU-based infrastructures.
The software stack comprises QuEST~4.2.0, Open MPI~5.1.0a1, PRRTE~5.0.0a1, OpenPMIX~7.0.0a1, and DMR v2.2.0.
All experiments scale up to 64 nodes, corresponding to the maximum node count available on our testbed.

To evaluate the two optimization decisions under controlled communication patterns, we have designed eight \emph{synthetic circuits}:
\begin{itemize}
    \item \emph{1D chain}, featuring only nearest-neighbour coupling and thus a strictly local interaction graph;
    \item \emph{QAOA-sparse}, a random MaxCut instance whose interactions are sparse but not local;
    \item \emph{Draper adder}, which incorporates a communication-heavy QFT block;
    \item \emph{Cuccaro adder}, which, in contrast to the Draper adder, relies on a mostly local carry chain;
    \item \emph{Ising Trotter}, a compute-heavy circuit employing nearest-neighbour coupling;
    \item \emph{SWAP routing}, which exercises the two-target cut model;
    \item \emph{QFT-epcc}, a control circuit where every qubit carries identical weight, yielding a cut of $k$ under any layout~\cite{adamski_energy_2023};
    \item \emph{Comm cliff}, a control circuit whose all--to--all core stays entirely
      local up to a fixed node count, beyond which part of the communication
      must go global.
\end{itemize}
Together, these circuits span a range from fully reducible to structurally irreducible communication patterns.

To verify that the proposed placement and node-count strategies generalize to unmodified, real-world workloads, we additionally evaluate circuits from the QASMBench's~\cite{li_qasmbench_2023} \emph{Large-scale} category\footnote{\url{https://github.com/pnnl/QASMBench\#large-scale}}. 
These are loaded through the OpenQASM~2.0 reader detailed in Section~\ref{subsec:impl-qasm}. 
Although the \emph{Large-scale} category encompasses circuits ranging from 28 to thousands of qubits, we restrict our evaluation to the 28--34 qubit range, which aligns with the per-node memory constraints of our cluster. 
Circuits relying on unsupported QASM constructs---such as custom gate definitions, \texttt{reset} instructions, or classical conditionals---are excluded from the benchmark corpus.

\subsection{Qubit Layout Optimization Study}

\subsubsection{Analytical Results}\label{subsec:eval-analytical}
Prior to running the circuits, we conduct an analytical evaluation to assess the impact of the global qubit placement optimizer. This analysis is performed on both synthetic and QASMBench circuits. Rather than allocating a state vector, we evaluate each circuit directly at a fixed node count. We apply a simulated-annealing search to determine the optimized qubit layout and report the communication cut (Equation~\ref{eq:communication-cut}) of the placements induced by two layouts with $k$ global qubits: the original \emph{identity} layout, denoted $cut_{id}(k)$, and the \emph{optimized} layout, whose cut is the lowest found by the search and is denoted $cut_{\min}(k)$ as in Equation~\ref{eq:min-communication-cut}.


``Synthetic'' rows in Table~\ref{tab:analytical} present the analytical communication cuts, comparing the identity and optimized layouts across all eight synthetic circuits at $N=30$ and $N=32$ qubits, distributed over 2 and 16 nodes. 
The range is defined excluding 1-node because it lacks internode communication; and up to $16$ nodes, since it effectively captures the trend observed at higher node counts---any increase in communication cut with scaling consistently manifests at $16$ nodes, mirroring the behavior at $32$ and $64$ nodes. 

\begin{table}[tbp]
\centering
\caption{Communication cut comparison between identity and optimized layouts. Unlike the other synthetic circuits, only the identity \emph{QAOA-sparse} with 16 nodes presents differences between 30- and 32-qubit circuits.}
\vspace{-0.25cm}
\label{tab:analytical}
\scriptsize
{\renewcommand{\arraystretch}{0.85}%
\setlength{\tabcolsep}{4pt}%
\begin{tabular}{@{}c@{\hspace{3pt}}l | c c c c}
\toprule
 & Circuit & \multicolumn{2}{c}{2 Nodes ($k=1$)} & \multicolumn{2}{c}{16 Nodes ($k=4$)} \\
\cmidrule(r){3-4} \cmidrule(l){5-6}
 & & $cut_{id}$ & $cut_{\min}$ & $cut_{id}$ & $cut_{\min}$ \\
\midrule
\multirow{8}{*}{\rotatebox[origin=c]{90}{Synthetic}}
 & QAOA-sparse    & 22   & 4  & 76--94 & 16 \\
 & Draper adder   & 2    & 0  & 8      & 0  \\
 & 1D chain       & 4    & 0  & 16     & 12 \\
 & QFT-epcc       & 1    & 1  & 4      & 4  \\
 & Comm cliff     & 0    & 0  & 0      & 0  \\
 & Cuccaro adder  & 1    & 1  & 10     & 8  \\
 & Ising Trotter  & 13   & 5  & 52     & 44 \\
 & SWAP routing   & 1    & 1  & 4      & 2  \\
\midrule
\multirow{12}{*}{\rotatebox[origin=c]{90}{QASMBench}}
 & qft\_n29               & 1     & 1     & 16     & 16     \\
 & qft\_n29\_transpiled   & 1     & 1     & 16     & 16     \\
 & adder\_n28             & 1     & 1     & 10     & 7      \\
 & adder\_n28\_transpiled & 1     & 1     & 10     & 10     \\
 & 32                     & 176   & 164   & 707    & 671    \\
 & bv\_n30                & 20    & 2     & 26     & 8      \\
 & bv\_n30\_transpiled    & 19    & 2     & 25     & 8      \\
 & dnn\_n33\_transpiled   & 8     & 2     & 32     & 25     \\
 & ising\_n34             & 5     & 3     & 20     & 18     \\
 & ising\_n34\_transpiled & 3     & 1     & 12     & 10     \\
 & knn\_n31\_transpiled   & 8     & 2     & 32     & 20     \\
 & vqe\_uccsd\_n28        & 4,152 & 4,152 & 28,392 & 20,664 \\
\bottomrule
\end{tabular}}
\vspace{-6pt}
\end{table}

Seven of the synthetic circuits exhibit identical cut reduction patterns at $N=30$ and $N=32$. This consistency arises because their fixed, hand-designed interaction topologies are invariant to the total qubit count; their communication profile depends solely on $k$. 
\emph{QAOA-sparse} serves as an exception, since its interaction graph is randomly instantiated across all $N$ qubits. 
Consequently, its identity cut at $16$ nodes increases from $76$ to $94$ when scaling from $N=30$ to $N=32$, resulting in a higher reduction ratio (from $78.9\%$ to $83.0\%$) despite the optimized cut remaining constant at $16$. 
The \emph{Draper adder} achieves $cut_{\min}(k)=0$ at both node counts, indicating complete elimination of inter-node communication; \emph{1D chain} reaches the same at $k=1$ but only partially reduces at $k=4$ ($16 \rightarrow 12$). The control circuits \emph{QFT-epcc} and \emph{Comm cliff} show no improvement from layout optimization at either node count. The \emph{Cuccaro adder} demonstrates optimization gains exclusively in the $16$-node configuration. 
Conversely, the \emph{Ising Trotter} circuit reduces exactly $8$ communication events at both node counts ($13 \rightarrow 5$ for $k=1$ and $52 \rightarrow 44$ for $k=4$). 
Its overall cut values decrease because the baseline identity cut grows with $k$, amplifying the relative impact of the optimization.

Similarly, we analytically evaluate the qubit layout optimization for the QASMBench circuits in the ``QASMBench'' rows in Table~\ref{tab:analytical}, applying the same node distributions (2 and 16). 
Circuits such as \emph{qft\_n29} and its transpiled variant (\emph{qft\_n29\_transpiled}) exhibit no improvement. 
The quantum Fourier transform (QFT) couples every qubit to all others via controlled-phase gates (decomposed into two CNOTs per gate) and applies a Hadamard gate to each qubit. 
Consequently, as with \emph{QFT-epcc}, every qubit carries an identical communication weight, rendering any permutation ineffective at reducing the cut. 
The \emph{adder\_n28} family also shows no improvement when $k=1$; at $k=4$, it benefits from the optimized layout, whereas its transpiled counterpart (\emph{adder\_n28\_transpiled}) does not. 
This mirrors the behavior of the \emph{Comm cliff} control in the synthetic benchmarks: when every qubit is involved nearly uniformly, layout optimization yields negligible gains.
The remaining circuits demonstrate consistent communication reductions across both node configurations.

\subsubsection{Performance Analysis}\label{subsec:eval-layout}
Unlike the analytical results, which only estimate communication through the cut, these experiments run the quantum circuit simulations and measure their execution times. 
The goal is to determine whether reducing the cut translates into lower runtime and to what extent the optimized layout improves performance over the identity layout.
A strong-scaling sweep runs each circuit at every node count from 2 to 64, skipping any node count whose per-node memory would exceed the cluster budget and, for the QASMBench set, any node count with $k\geq N$ or outside the circuit's native qubit window.

The results have been validated (final states are compared before and after the placement optimization) and the times correspond to the mean of four repetitions.
Run--to--run variation across them is negligible on the synthetic set, averaging $0.14\%$ of the mean (both $N=30$ and $N=32$) and never exceeding $1\%$.
On QASMBench, the average is $0.17\%$, reaching $1.87\%$ in \emph{knn\_n31\_transpiled} at 16 nodes.

``Synthetic'' rows in Table~\ref{tab:timing-n16} report the measurements on the eight synthetic circuits at 16 nodes ($k=4$), $N=30$ and $N=32$, the mean wall time per repetition without and with the layout optimization, and the resulting speedup.

The two controls, \emph{QFT-epcc} and \emph{Comm cliff}, sit at $1.00\times$ at both sizes, confirming the layout does not worsen a circuit it cannot help. 
Every other circuit gains at both sizes, from $1.13$--$1.14\times$ (\emph{Ising Trotter}, the most compute-heavy of the set) up to $1.87\times$ at $N=30$ and $2.09\times$ at $N=32$ (\emph{QAOA-sparse}, the sparsest interaction pattern): its identity cut grows with $N$, so the layout has more communication to remove at the larger size, consistent with the graph structure described in Section~\ref{subsec:placement}.

\begin{table}[tbp]
\centering
\caption{Timing at 16 nodes ($k=4$).}
\vspace{-0.25cm}
\label{tab:timing-n16}
\scriptsize
{\renewcommand{\arraystretch}{0.85}%
\setlength{\tabcolsep}{4pt}%
\begin{tabular}{@{}c@{\hspace{3pt}}llrrr}
\toprule
 & Circuit & $N$ & $T_{\text{nolayout}}$ (s) & $T_{\text{layout}}$ (s) & Speedup \\
\midrule
\multirow{16}{*}{\rotatebox[origin=c]{90}{Synthetic}}
 & QAOA-sparse   & 30 & 6.63  & 3.55  & 1.87$\times$ \\
 &               & 32 & 34.93 & 16.73 & 2.09$\times$ \\
 & Draper adder  & 30 & 2.40  & 1.57  & 1.53$\times$ \\
 &               & 32 & 11.77 & 8.09  & 1.45$\times$ \\
 & 1D chain      & 30 & 1.46  & 1.09  & 1.34$\times$ \\
 &               & 32 & 6.39  & 5.00  & 1.28$\times$ \\
 & QFT-epcc      & 30 & 1.83  & 1.83  & 1.00$\times$ \\
 &               & 32 & 9.59  & 9.59  & 1.00$\times$ \\
 & Comm cliff    & 30 & 3.02  & 3.02  & 1.00$\times$ \\
 &               & 32 & 16.25 & 16.28 & 1.00$\times$ \\
 & Cuccaro adder & 30 & 0.92  & 0.73  & 1.25$\times$ \\
 &               & 32 & 4.38  & 3.31  & 1.32$\times$ \\
 & Ising Trotter & 30 & 7.01  & 6.22  & 1.13$\times$ \\
 &               & 32 & 33.00 & 28.91 & 1.14$\times$ \\
 & SWAP routing  & 30 & 0.23  & 0.18  & 1.26$\times$ \\
 &               & 32 & 1.12  & 0.90  & 1.25$\times$ \\
\midrule
\multirow{12}{*}{\rotatebox[origin=c]{90}{QASMBench}}
 & qft\_n29               & 29 & 79.37    & 79.34    & 1.00$\times$ \\
 & qft\_n29\_transpiled   & 29 & 82.03    & 82.07    & 1.00$\times$ \\
 & adder\_n28             & 28 & 1.31     & 1.24     & 1.05$\times$ \\
 & adder\_n28\_transpiled & 28 & 9.72     & 9.71     & 1.00$\times$ \\
 & 32                     & 32 & 536.11   & 529.20   & 1.01$\times$ \\
 & bv\_n30                & 30 & 2.10     & 1.15     & 1.83$\times$ \\
 & bv\_n30\_transpiled    & 30 & 2.58     & 1.74     & 1.48$\times$ \\
 & dnn\_n33\_transpiled   & 33 & 56.32    & 52.63    & 1.07$\times$ \\
 & ising\_n34             & 34 & 58.53    & 58.27    & 1.00$\times$ \\
 & ising\_n34\_transpiled & 34 & 34.16    & 31.85    & 1.07$\times$ \\
 & knn\_n31\_transpiled   & 31 & 9.26     & 7.70     & 1.20$\times$ \\
 & vqe\_uccsd\_n28        & 28 & 10763.39 & 10788.76 & 1.00$\times$ \\
\bottomrule
\end{tabular}}
\vspace{-6pt}
\end{table}

In more detail, ``Synthetic'' rows in Table~\ref{tab:speedup-vs-k} extend the speedup results for every node count. 
At $N=30$ the benefit grows with $k$ up to 16 nodes, peaking at $1.87\times$ for \emph{QAOA-sparse} and $1.53\times$ for \emph{Draper adder}, then collapses toward $1.00\times$ at 32 and 64 nodes: past the distribution floor ($N-k<26$), QuEST replicates the state vector instead of distributing it, nothing communicates, and the layout has nothing left to save. 
At $N=32$ the floor falls at $k=7$ (128 nodes), beyond both the measured range and the 64-node testbed itself, so the benefit keeps growing instead, reaching $2.27\times$ for \emph{QAOA-sparse} and $1.65\times$ for \emph{Draper adder} at 64 nodes. 
The two controls, \emph{QFT-epcc} and \emph{Comm cliff}, stay at 1.00× throughout at both sizes, as neither had any communication for the layout to eliminate.

\begin{table}[tbp]
\centering
\caption{Layout speedup per node count.}
\vspace{-0.25cm}
\label{tab:speedup-vs-k}
\scriptsize
{\renewcommand{\arraystretch}{0.85}%
\setlength{\tabcolsep}{4pt}%
\begin{tabular}{@{}c@{\hspace{3pt}}ll|rrrrrr}
\toprule
 & & & \multicolumn{6}{c}{Node Count} \\
 & Circuit & $N$ & 2 & 4 & 8 & 16 & 32 & 64 \\
\midrule
\multirow{16}{*}{\rotatebox[origin=c]{90}{Synthetic}}
 & QAOA-sparse    & 30 & \cellcolor[HTML]{AFC7E4}1.20$\times$ & \cellcolor[HTML]{97B2D2}1.36$\times$ & \cellcolor[HTML]{7694BA}1.58$\times$ & \cellcolor[HTML]{4A6D99}\color{white}1.87$\times$ & \cellcolor[HTML]{CCE1FA}1.01$\times$ & \cellcolor[HTML]{C8DEF8}1.03$\times$ \\
 &                & 32 & \cellcolor[HTML]{AAC3E1}1.23$\times$ & \cellcolor[HTML]{809DC2}1.51$\times$ & \cellcolor[HTML]{6C8BB2}1.65$\times$ & \cellcolor[HTML]{2A5080}\color{white}2.09$\times$ & \cellcolor[HTML]{20477A}\color{white}2.15$\times$ & \cellcolor[HTML]{0E376C}\color{white}2.27$\times$ \\
 & Draper adder   & 30 & \cellcolor[HTML]{C4DAF4}1.06$\times$ & \cellcolor[HTML]{B6CEEA}1.15$\times$ & \cellcolor[HTML]{A0BAD9}1.30$\times$ & \cellcolor[HTML]{7D9BBF}1.53$\times$ & \cellcolor[HTML]{CDE2FB}1.00$\times$ & \cellcolor[HTML]{C4DAF4}1.06$\times$ \\
 &                & 32 & \cellcolor[HTML]{C6DBF5}1.05$\times$ & \cellcolor[HTML]{B2CAE7}1.18$\times$ & \cellcolor[HTML]{AFC7E4}1.20$\times$ & \cellcolor[HTML]{8AA6C8}1.45$\times$ & \cellcolor[HTML]{85A2C5}1.48$\times$ & \cellcolor[HTML]{6C8BB2}1.65$\times$ \\
 & 1D chain       & 30 & \cellcolor[HTML]{ACC4E2}1.22$\times$ & \cellcolor[HTML]{B4CBE8}1.17$\times$ & \cellcolor[HTML]{A9C2E0}1.24$\times$ & \cellcolor[HTML]{9AB4D5}1.34$\times$ & \cellcolor[HTML]{CDE2FB}0.98$\times$ & \cellcolor[HTML]{CDE2FB}0.98$\times$ \\
 &                & 32 & \cellcolor[HTML]{A9C2E0}1.24$\times$ & \cellcolor[HTML]{A9C2E0}1.24$\times$ & \cellcolor[HTML]{A3BCDC}1.28$\times$ & \cellcolor[HTML]{A3BCDC}1.28$\times$ & \cellcolor[HTML]{A9C2E0}1.24$\times$ & \cellcolor[HTML]{A3BCDC}1.28$\times$ \\
 & QFT-epcc       & 30 & \cellcolor[HTML]{CDE2FB}1.00$\times$ & \cellcolor[HTML]{CDE2FB}1.00$\times$ & \cellcolor[HTML]{CDE2FB}1.00$\times$ & \cellcolor[HTML]{CDE2FB}1.00$\times$ & \cellcolor[HTML]{CDE2FB}1.00$\times$ & \cellcolor[HTML]{CDE2FB}1.00$\times$ \\
 &                & 32 & \cellcolor[HTML]{CDE2FB}1.00$\times$ & \cellcolor[HTML]{CDE2FB}1.00$\times$ & \cellcolor[HTML]{CDE2FB}1.00$\times$ & \cellcolor[HTML]{CDE2FB}1.00$\times$ & \cellcolor[HTML]{CDE2FB}1.00$\times$ & \cellcolor[HTML]{CDE2FB}1.00$\times$ \\
 & Comm cliff     & 30 & \cellcolor[HTML]{CDE2FB}1.00$\times$ & \cellcolor[HTML]{CDE2FB}1.00$\times$ & \cellcolor[HTML]{CDE2FB}1.00$\times$ & \cellcolor[HTML]{CDE2FB}1.00$\times$ & \cellcolor[HTML]{CDE2FB}1.00$\times$ & \cellcolor[HTML]{CDE2FB}1.00$\times$ \\
 &                & 32 & \cellcolor[HTML]{CDE2FB}1.00$\times$ & \cellcolor[HTML]{CDE2FB}1.00$\times$ & \cellcolor[HTML]{CDE2FB}1.00$\times$ & \cellcolor[HTML]{CDE2FB}1.00$\times$ & \cellcolor[HTML]{CDE2FB}1.00$\times$ & \cellcolor[HTML]{CDE2FB}1.00$\times$ \\
 & Cuccaro adder  & 30 & \cellcolor[HTML]{CDE2FB}1.00$\times$ & \cellcolor[HTML]{B6CEEA}1.15$\times$ & \cellcolor[HTML]{98B3D4}1.35$\times$ & \cellcolor[HTML]{A8C0DF}1.25$\times$ & \cellcolor[HTML]{CADFF9}1.02$\times$ & \cellcolor[HTML]{C6DBF5}1.05$\times$ \\
 &                & 32 & \cellcolor[HTML]{CDE2FB}1.00$\times$ & \cellcolor[HTML]{A9C2E0}1.24$\times$ & \cellcolor[HTML]{A4BEDD}1.27$\times$ & \cellcolor[HTML]{9DB7D7}1.32$\times$ & \cellcolor[HTML]{AFC7E4}1.20$\times$ & \cellcolor[HTML]{B0C8E6}1.19$\times$ \\
 & Ising Trotter  & 30 & \cellcolor[HTML]{C6DBF5}1.05$\times$ & \cellcolor[HTML]{C6DBF5}1.05$\times$ & \cellcolor[HTML]{C2D9F3}1.07$\times$ & \cellcolor[HTML]{BAD1EC}1.13$\times$ & \cellcolor[HTML]{CDE2FB}1.00$\times$ & \cellcolor[HTML]{CDE2FB}1.00$\times$ \\
 &                & 32 & \cellcolor[HTML]{C4DAF4}1.06$\times$ & \cellcolor[HTML]{C1D7F2}1.08$\times$ & \cellcolor[HTML]{C0D6F1}1.09$\times$ & \cellcolor[HTML]{B8CFEB}1.14$\times$ & \cellcolor[HTML]{C0D6F1}1.09$\times$ & \cellcolor[HTML]{C0D6F1}1.09$\times$ \\
 & SWAP routing   & 30 & \cellcolor[HTML]{CDE2FB}1.00$\times$ & \cellcolor[HTML]{C4DAF4}1.06$\times$ & \cellcolor[HTML]{BAD1EC}1.13$\times$ & \cellcolor[HTML]{A6BFDE}1.26$\times$ & \cellcolor[HTML]{C7DDF6}1.04$\times$ & \cellcolor[HTML]{C7DDF6}1.04$\times$ \\
 &                & 32 & \cellcolor[HTML]{CCE1FA}1.01$\times$ & \cellcolor[HTML]{88A4C7}1.46$\times$ & \cellcolor[HTML]{C7DDF6}1.04$\times$ & \cellcolor[HTML]{A8C0DF}1.25$\times$ & \cellcolor[HTML]{AFC7E4}1.20$\times$ & \cellcolor[HTML]{AEC6E3}1.21$\times$ \\
\midrule
\multirow{12}{*}{\rotatebox[origin=c]{90}{QASMBench}}
 & qft\_n29               & 29 & \cellcolor[HTML]{CDE2FB}1.00$\times$ & \cellcolor[HTML]{CDE2FB}1.00$\times$ & \cellcolor[HTML]{CDE2FB}1.00$\times$ & \cellcolor[HTML]{CDE2FB}1.00$\times$ & \cellcolor[HTML]{CDE2FB}1.00$\times$ & \cellcolor[HTML]{CDE2FB}1.00$\times$ \\
 & qft\_n29\_transpiled   & 29 & \cellcolor[HTML]{CDE2FB}1.00$\times$ & \cellcolor[HTML]{CDE2FB}1.00$\times$ & \cellcolor[HTML]{CDE2FB}1.00$\times$ & \cellcolor[HTML]{CDE2FB}1.00$\times$ & \cellcolor[HTML]{CDE2FB}1.00$\times$ & \cellcolor[HTML]{CDE2FB}1.00$\times$ \\
 & adder\_n28             & 28 & \cellcolor[HTML]{CCE1FA}1.01$\times$ & \cellcolor[HTML]{C1D7F2}1.08$\times$ & \cellcolor[HTML]{CADFF9}1.02$\times$ & \cellcolor[HTML]{C6DBF5}1.05$\times$ & \cellcolor[HTML]{C0D6F1}1.09$\times$ & \cellcolor[HTML]{C4DAF4}1.06$\times$ \\
 & adder\_n28\_transpiled & 28 & \cellcolor[HTML]{CDE2FB}1.00$\times$ & \cellcolor[HTML]{CDE2FB}1.00$\times$ & \cellcolor[HTML]{CDE2FB}1.00$\times$ & \cellcolor[HTML]{CDE2FB}1.00$\times$ & \cellcolor[HTML]{CDE2FB}1.00$\times$ & \cellcolor[HTML]{CDE2FB}1.00$\times$ \\
 & 32                     & 32 & \cellcolor[HTML]{CCE1FA}1.01$\times$ & \cellcolor[HTML]{CCE1FA}1.01$\times$ & \cellcolor[HTML]{CCE1FA}1.01$\times$ & \cellcolor[HTML]{CCE1FA}1.01$\times$ & \cellcolor[HTML]{CADFF9}1.02$\times$ & \cellcolor[HTML]{CADFF9}1.02$\times$ \\
 & bv\_n30                & 30 & \cellcolor[HTML]{0D366B}\color{white}2.28$\times$ & \cellcolor[HTML]{2C5283}\color{white}2.07$\times$ & \cellcolor[HTML]{2B5181}\color{white}2.08$\times$ & \cellcolor[HTML]{50729E}\color{white}1.83$\times$ & \cellcolor[HTML]{C6DBF5}1.05$\times$ & \cellcolor[HTML]{C6DBF5}1.05$\times$ \\
 & bv\_n30\_transpiled    & 30 & \cellcolor[HTML]{7F9CC0}1.52$\times$ & \cellcolor[HTML]{84A0C4}1.49$\times$ & \cellcolor[HTML]{7694BA}1.58$\times$ & \cellcolor[HTML]{85A2C5}1.48$\times$ & \cellcolor[HTML]{CADFF9}1.02$\times$ & \cellcolor[HTML]{CADFF9}1.02$\times$ \\
 & dnn\_n33\_transpiled   & 33 & & \cellcolor[HTML]{C4DAF4}1.06$\times$ & \cellcolor[HTML]{C2D9F3}1.07$\times$ & \cellcolor[HTML]{C2D9F3}1.07$\times$ & \cellcolor[HTML]{BCD3EF}1.11$\times$ & \cellcolor[HTML]{BED5F0}1.10$\times$ \\
 & ising\_n34             & 34 & & & \cellcolor[HTML]{C7DDF6}1.04$\times$ & \cellcolor[HTML]{CDE2FB}1.00$\times$ & \cellcolor[HTML]{C2D9F3}1.07$\times$ & \cellcolor[HTML]{C7DDF6}1.04$\times$ \\
 & ising\_n34\_transpiled & 34 & & & \cellcolor[HTML]{C4DAF4}1.06$\times$ & \cellcolor[HTML]{C2D9F3}1.07$\times$ & \cellcolor[HTML]{BAD1EC}1.13$\times$ & \cellcolor[HTML]{B6CEEA}1.15$\times$ \\
 & knn\_n31\_transpiled   & 31 & \cellcolor[HTML]{BBD2EE}1.12$\times$ & \cellcolor[HTML]{BAD1EC}1.13$\times$ & \cellcolor[HTML]{B0C8E6}1.19$\times$ & \cellcolor[HTML]{AFC7E4}1.20$\times$ & \cellcolor[HTML]{AFC7E4}1.20$\times$ & \cellcolor[HTML]{C1D7F2}1.08$\times$ \\
  & vqe\_uccsd\_n28        & 28 & \cellcolor[HTML]{CDE2FB}1.00$\times$ & \cellcolor[HTML]{C4DAF4}1.06$\times$ & \cellcolor[HTML]{CDE2FB}0.99$\times$ & \cellcolor[HTML]{CDE2FB}1.00$\times$ & \cellcolor[HTML]{CADFF9}1.02$\times$ & \cellcolor[HTML]{CDE2FB}1.00$\times$ \\
\bottomrule
\end{tabular}}
\vspace{-6pt}
\end{table}

``QASMBench'' rows in Table~\ref{tab:timing-n16} report the same timing-mode result as in ``Synthetic'' rows but for the QASMBench \emph{Large-scale} circuits, at 16 nodes ($k=4$): the mean wall time per iteration without and with the layout, and the resulting speedup.

As with the synthetic set, ``QASMBench'' rows in Table~\ref{tab:speedup-vs-k} present the times and speedups but for the QASMBench \emph{Large-scale} circuits (blank where a node count does not fit the circuit in memory). 
\emph{bv\_n30} and \emph{bv\_n30\_transp.} stay large and comparatively stable up to 8 nodes before the same collapse past 16, and \emph{knn\_n31\_transp.} keeps growing up to 32 nodes before falling back at 64.
The QFT-like circuits, \emph{qft\_n29} and \emph{qft\_n29\_transp.}, are flat at $1.00\times$ throughout, above or beyond the floor, since the layout can never reduce their (nonzero, irreducible) communication.

The circuit with the largest analytical reduction in communication at two nodes, \emph{bv\_n30}, also achieves the highest measured speedup at that node count ($2.28\times$). 
At 16 nodes, however, the relationship between cut reduction and speedup is less direct. 
The time saved by eliminating a communicating gate depends not only on the number of removed gates, but also on the size of the local state-vector block exchanged for each gate, which is proportional to \(2^{N-k}\) (Section~\ref{subsec:resources}).
For example, \emph{adder\_n28} has the largest cut reduction among the circuits that benefit from optimization at \(k=4\) ($30\%$), but achieves the smallest speedup ($1.05\times$). 
By comparison, \emph{dnn\_n33\_transpiled} and \emph{ising\_n34\_transp.} reduce their cuts by only $21.9\%$ and $16.7\%$, respectively, yet achieve speedups of $1.07\times$ in both cases. 
This difference arises because, at the same \(k\), the 28-qubit circuit exchanges local blocks that are \(2^{5}\)--\(2^{6}\) times smaller than those of the 33- and 34-qubit circuits. 
Consequently, removing a communicating gate saves proportionally less time for \emph{adder\_n28}. 
Finally, \emph{qft\_n29} and \emph{qft\_n29\_transp.} show no speedup: their communication cuts are unchanged by placement at every \(k\) (Section~\ref{subsec:eval-analytical}), so both remain at \(1.00\times\), consistent with the analytical results.

\subsection{Node-count Efficiency Study}\label{subsec:eval-efficiency-model}
The placement optimizer selects an efficient qubit layout for a given number of nodes, but the node count itself must also be determined. 
Allocating more nodes can increase available memory and parallelism; however, it also increases the number of global qubits and may introduce additional communication and synchronization overhead. 
As a result, the configuration with the largest allocation is not necessarily the fastest, nor is the fastest configuration necessarily the most resource-efficient.

\begin{figure}[tbp]
\centering
\includegraphics[width=0.75\columnwidth]{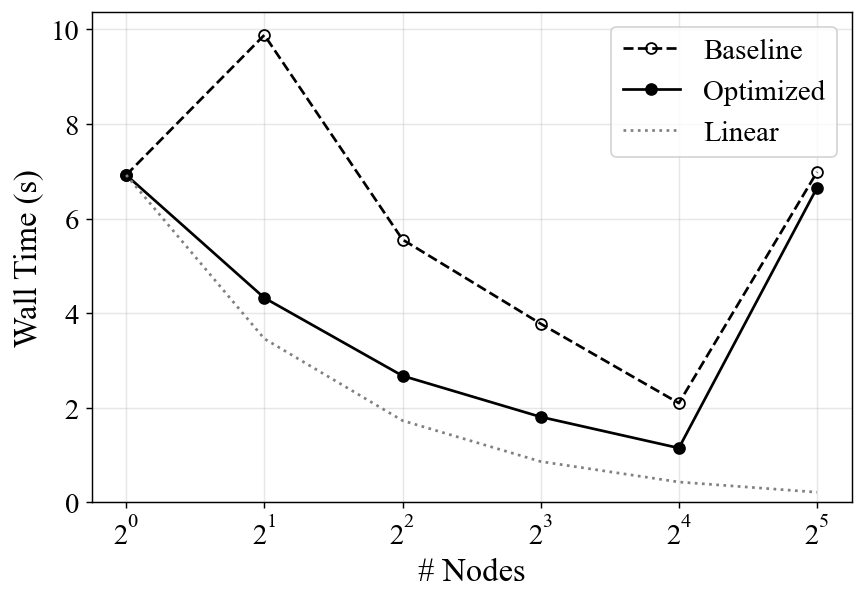}
\vspace{-0.25cm}
\caption{\emph{bv\_n30} wall time vs.\ node count, without and with the qubit layout optimization. Dotted line marks linear scaling.}
\label{fig:nodecount-bv30}
\Description{Line chart of wall time in seconds against node count from 1 to 32 for bv\_n30. The baseline peaks near 10 s at 2 nodes and then falls to about 2 s at 16 nodes. The optimized layout decreases steadily from about 7 s to about 1.2 s at 16 nodes, staying closer to the linear-scaling line. Both rise to about 7 s at 32 nodes.}
\vspace{-6pt}
\end{figure}

We begin with an example that illustrates the effect of scaling and the extent to which layout optimization mitigates the overhead of using additional nodes in distributed execution. 
Figure~\ref{fig:nodecount-bv30} shows the execution time of \emph{bv\_n30} for different node counts. 
The optimized configuration remains consistently closer to the linear-scaling trend than the identity layout, which serves as the baseline.

At 32 nodes, $N-k$ falls below QuEST's distribution floor.
Beyond this point, the state vector cannot be divided into smaller local portions, so additional nodes do not expose more computational work. 

We next validate the model introduced in Section~\ref{subsec:resources}.
The model was calibrated on MN5, as described in Section~\ref{subsec:impl-resources}, and predicts $O_{k_{\min}}(k)$, namely, the deviation from linear scaling relative to the smallest feasible node count for each circuit. 
This reference configuration corresponds to one node, except for the circuits whose memory requirements prevent execution on a single node.

The prediction uses only static information: the circuit and its optimized layout. 
It therefore requires neither a preliminary execution nor additional runtime measurements. 
To assess its accuracy, we compare the predicted performance curves with the wall times measured in the previous subsection.

\begin{figure*}[tbp]
\centering
\includegraphics[width=0.85\textwidth]{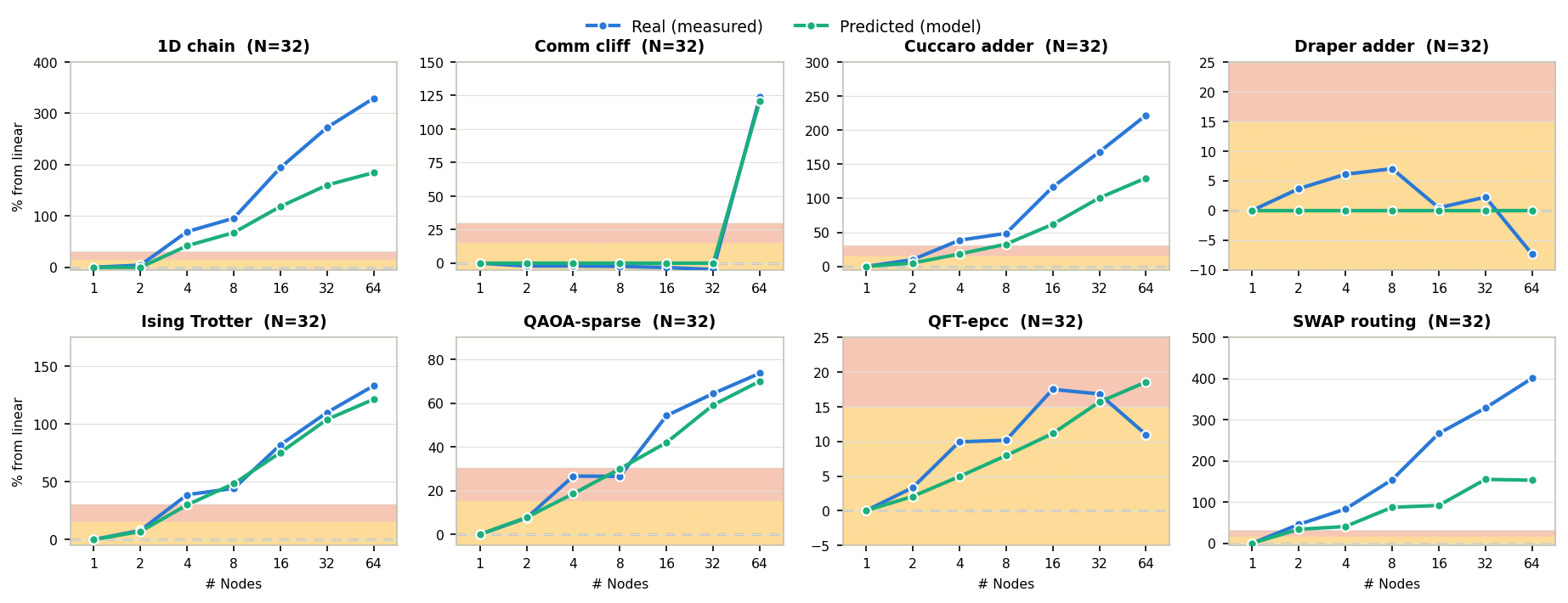}
\vspace{-0.25cm}
\caption{Predicted vs.\ measured deviation from linear scaling, eight synthetic circuits, $N=32$.}
\label{fig:deviation-synthetic}
\Description{Grid of eight line charts, one per synthetic circuit with 32 qubits: 1D chain, Comm cliff, Cuccaro adder, Draper adder, Ising Trotter, QAOA-sparse, QFT-epcc and SWAP routing. Each chart plots the percentage deviation from linear scaling against node count from 1 to 64, with a measured curve and a model-predicted curve, over shaded bands marking the 15\% and 30\% degradation thresholds.}
\vspace{-12pt}
\end{figure*}

Figure~\ref{fig:deviation-synthetic} compares the predicted and measured degradation curves for the eight synthetic circuits with $N=32$. 
We select this circuit size because it offers the broadest feasible range above QuEST's distribution floor, namely $k\in[0,6]$, corresponding to node counts of up to 64. 
The feasible ranges are narrower for $N=30$ and $N=28$, with $k\in[0,4]$ and $k\in[0,2]$, respectively. 
The smaller circuits ($N=16$, $N=22$, and $N=26$) reach the distribution floor almost immediately and consequently provide too few distributed configurations for a meaningful validation. 

\begin{figure*}[tbp]
\centering
\includegraphics[width=0.85\textwidth]
{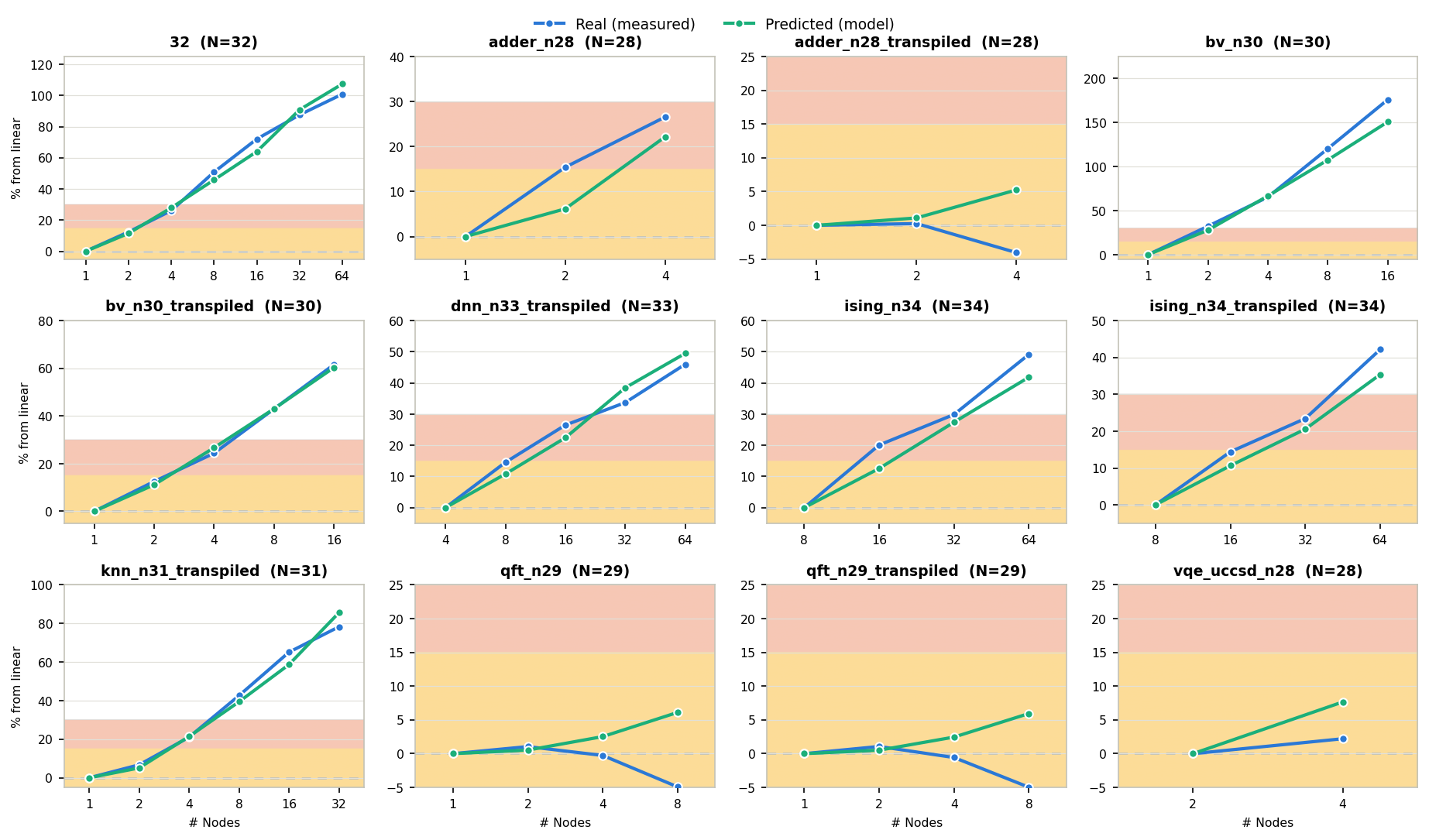}
\vspace{-0.25cm}
\caption{Predicted vs.\ measured deviation from linear scaling of the QASMBench \emph{Large-scale} circuits.}
\label{fig:deviation-qasmbench}
\Description{Grid of twelve line charts, one per QASMBench large-scale circuit, including adder\_n28, bv\_n30, dnn\_n33, ising\_n34, knn\_n31, qft\_n29 and vqe\_uccsd\_n28 and their transpiled variants. Each chart plots the percentage deviation from linear scaling against node count, with a measured curve and a model-predicted curve, over shaded bands marking the 15\% and 30\% degradation thresholds.}
\vspace{-6pt}
\end{figure*}

Figure~\ref{fig:deviation-qasmbench} shows the corresponding comparison for the QASMBench \emph{Large-scale} circuits. 
Both figures also show two degradation-threshold regions using colored bands: $\tau=30\%$ and $\tau=15\%$, represented in red and yellow colors, respectively. 
The first threshold follows the recommendation of the MN5 support team, which considers communication efficiencies above 70\% desirable and therefore corresponds to a maximum degradation of $\tau=30\%$ (see Section~\ref{subsec:resources}). 
We additionally evaluate the stricter threshold $\tau=15\%$, which limits the permitted degradation to half of that value.

For each threshold, we classify a prediction as correct when the node count $k_{\text{eff}}$ selected from the predicted curve is the same as the node count selected from the measured curve. 
Table~\ref{tab:eff-validation} summarizes the resulting accuracy for both circuit sets. 
When the predicted and measured curves lead to the same threshold band, the selected node count is, by construction, the most efficient feasible choice under that threshold: $k_{\text{eff}}$ is defined as the largest feasible value of $k$ satisfying the degradation constraint. 

\begin{table}[t]
\centering
\caption{Node-count efficiency model: agreement between predicted and measured degradation.}
\vspace{-0.25cm}
\label{tab:eff-validation}
\scriptsize
{\renewcommand{\arraystretch}{0.85}%
\setlength{\tabcolsep}{4pt}%
\begin{tabular}{lc|cc}
\toprule
Corpus & Circuits & $\tau$ & Accuracy \\
\midrule
Synthetic ($N=32$)          & 8  & 15\% & 88\% \\
                             &    & 30\% & 88\% \\
QASMBench \emph{Large-scale} & 12 & 15\% & 83\% \\
                             &    & 30\% & 92\% \\
\bottomrule
\end{tabular}}
\vspace{-6pt}
\end{table}

Figure~\ref{fig:deviation-synthetic} showcases the results for the synthetic circuits.
At $\tau=15\%$, \emph{QFT-epcc} is predicted to have a degradation of $11.1\%$, whereas the measured value is $17.5\%$ at 16 nodes.
At $\tau=30\%$, the \emph{Cuccaro adder} is predicted at $18.2\%$, but reaches $38.5\%$ at 4 nodes. 
In both cases, the prediction falls below the threshold while the measurement exceeds it, by $2.5$ and $8.5$ percentage points, respectively, so the model selects a larger node count than the measurements justify.
Neither circuit is misclassified at both thresholds: the \emph{Cuccaro adder} remains below $\tau=15\%$ at the selected node count, with a measured degradation of $10\%$ at 2 nodes, while \emph{QFT-epcc} remains below $\tau=30\%$ throughout the measured range. 
The discrepancies are therefore limited to one threshold--circuit pair for each circuit rather than reflecting a general failure to model either workload.

For the QASMBench circuits in Figure~\ref{fig:deviation-qasmbench}, \emph{bv\_n30} follows the model closely in the range relevant to the thresholds. 
At 2 and 4 nodes, the difference between predicted and measured degradation is approximately $1$--$5$ percentage points, although the degradation itself reaches $27$--$65\%$. 
At 8 and 16 nodes, the model underestimates degradation by approximately $13$--$25$ percentage points.
However, the measured degradation at these node counts is already between $107\%$ and $176\%$, well above both thresholds considered in this study.
These configurations would therefore not be selected for either $\tau=15\%$ or $\tau=30\%$. 
The larger discrepancy at high node counts does not affect the resource decisions made by the model, since those configurations are already outside the acceptable-performance range.

The \emph{qft\_n29} family exhibits a different behavior. 
Most of its gates are diagonal, so the total gate count $GA$ remains large relative to the communication workload $W_{\mathrm{comm}}(k)$. 
Within the feasible distributed range, the model correctly predicts the low degradation up to $k=3$, or 8 nodes, which is the distribution floor for $N=29$. 
At this point, the measured degradation is $-5.4\%$, compared with a predicted value of $6.1\%$.
Beyond the distribution floor, QuEST replicates the state vector instead of distributing additional computational work. 
The communication-based model does not apply in this regime, and the measured degradation consequently increases abruptly to $1,508\%$ at 16 nodes.

The \emph{adder\_n28} circuit follows the same pattern at a lower distribution floor. 
Its feasible range ends at $k=2$, corresponding to 4 nodes. 
At this point, the predicted and measured degradations are $22.2\%$ and $21.3\%$, respectively. 
Once the floor is exceeded, the measured degradation rises sharply to $693.5\%$ at 8 nodes. 
These two circuits therefore confirm that the model is accurate within the range in which QuEST continues to distribute the state vector, but should not be extrapolated beyond the distribution floor.

\subsection{Automatic Resource Adjustment}
\label{subsec:eval-automatic-adjustment}

\begin{table*}[tbp]
\centering
\scriptsize
\caption{Performance results of \emph{bv\_n30} ($N=30$) and \emph{vqe\_uccsd\_n28} ($N=28$).}
\vspace{-0.25cm}
\label{tab:eff-results}
{\renewcommand{\arraystretch}{0.85}%
\setlength{\tabcolsep}{4pt}%
\begin{tabular}{llrrrrrr}
\toprule
 & & & \multicolumn{2}{c}{Overhead} & & & \\
\cmidrule(lr){4-5}
Circuit & Strategy & \# Nodes & Optimization (s) & Reconfiguration (s) & Execution (s) & Wall time (s) & Cost (node-hour) \\
\midrule
\multirow{3}{*}{\emph{bv\_n30}}
 & Greedy (original circuit)             & 64        & N/A   & N/A   & 7.116 & 7.116 & 0.1265 \\
 & Performance-driven (original circuit) & 16        & N/A   & N/A   & 2.161 & 2.161 & 0.0096 \\
 & Automatic (optimized layout)          & 1 $\to$ 2 & 0.016 & 1.359 & 5.526 & 6.901 & 0.0035 \\
\midrule
\multirow{3}{*}{\emph{vqe\_uccsd\_n28}}
 & Greedy (original circuit)             & 64        & N/A   & N/A   & 10,813.00 & 10,813.00 & 192.23 \\
 & Performance-driven (original circuit) & 4         & N/A   & N/A   & 2,834.97  & 2,834.97  & 3.15   \\
 & Automatic (optimized layout)          & 1 $\to$ 4 & 6.27  & 2.79  & 2,765.55  & 2,774.61  & 3.07   \\
\bottomrule
\end{tabular}}
\vspace{-6pt}
\end{table*}

In production cluster environments, users must request their node allocation before launching a job. In practice, several strategies can be used to determine the amount of resources to allocate. For instance:
\begin{itemize}
    \item \emph{Conservative}: request the minimum number of nodes required to run the circuit. This minimizes resource usage but can lead to long execution times.

    \item \emph{Greedy}: request the maximum permitted number of nodes, or an unnecessarily large allocation. This is not only the most expensive strategy; for QuEST, it can also increase execution time. Once the distribution floor is exceeded, additional nodes replicate the state vector and introduce MPI overhead without exposing further computational parallelism.

    \item \emph{Performance-driven}: perform a preliminary sweep over the available node counts and select the configuration with the shortest execution time. This strategy is useful when the same circuit, or a sufficiently similar circuit, will be executed repeatedly, since the sweep itself incurs the cost that the selected configuration is intended to avoid.
\end{itemize}

Our approach automates this selection by identifying the largest node count whose predicted degradation remains below a user-defined threshold $\tau$. The selected configuration is determined before execution, while $\tau$ allows the user to control the trade-off between performance and resource efficiency.

We implement this mechanism using DMR, the dynamic resource-management framework introduced in Section~\ref{sec:background}. DMR resizes the job at runtime, so users do not need to know the final node count
when submitting it. Instead, the application is submitted with an initial user-specified allocation and is subsequently resized to the node count recommended by the model.

QuCSLO introduces additional overhead from layout optimization and
resource reconfiguration. We therefore evaluate whether the resulting resource savings and execution performance justify these costs using two representative circuits, \emph{bv\_n30} and \emph{vqe\_uccsd\_n28}. For each circuit, we compare automatic resource adjustment with execution without the framework, using a static node count selected according to the user-based strategies described above.

Table~\ref{tab:eff-results} illustrates this trade-off for
\emph{bv\_n30} ($N=30$). It compares two baseline configurations using the
original circuit, corresponding to the greedy and performance-driven
strategies, with an automatically configured execution that applies the
optimized layout and selects the node count using $\tau=30\%$.

For the automatic configuration, the model selects $k_{\text{eff}}=1$,
corresponding to two nodes, and applies the optimized qubit layout. This
allocation is substantially smaller than those used by the greedy and
performance-driven configurations, which use 64 and 16 nodes, respectively.
As shown in columns 4-5, layout optimization and reconfiguration from
one to two nodes introduce an overhead of 1.375~s. This accounts for
approximately 20\% of the total wall time of 6.901~s.

Compared with the performance-driven configuration, the automatically selected
configuration is $3.19\times$ slower but reduces node-hour cost by a factor of
$2.74$. Compared with the greedy configuration, it achieves a nearly identical
execution time, completing approximately 3\% faster while reducing
computational cost by a factor of $36$. These results illustrate the intended
trade-off: the model avoids excessive resource allocation while retaining
performance close to that of a substantially more expensive configuration.

Table~\ref{tab:eff-results} presents the corresponding results for
\emph{vqe\_uccsd\_n28} ($N=28$). In this case, the performance-driven and
automatically configured executions achieve approximately 74\% lower wall time
and 98\% lower node-hour cost than the greedy configuration. The difference is
explained by the fact that the greedy allocation of 64 nodes exceeds the
distribution floor of this circuit. For \emph{vqe\_uccsd\_n28}, the memory
limit is reached at $k_{\mathrm{mem}}=2$, corresponding to four nodes. 

The automatic and performance-driven configurations both use four nodes.
Nevertheless, the automatic configuration achieves lower execution time and
lower wall time, even after accounting for the 9.06~s overhead introduced by
layout optimization and resource reconfiguration. Unlike
\emph{bv\_n30}, the optimized layout changes very little for
\emph{vqe\_uccsd\_n28} at four nodes. Consequently, placement optimization
provides limited additional benefit, and the main advantage of the automatic
configuration is the elimination of the preliminary performance sweep and the
avoidance of excessive resource allocation.

\section{Discussion}\label{sec:discussion}
The current approach is subject to \textbf{limitations}:

1) Platform-specific communication model: The communication cost term $W_{\text{comm}}(k)$ is derived from QuEST's state-vector communication strategy. This tightly couples the model to the studied implementation; simulators employing alternative data-exchange patterns would require a revised communication framework.

2) Hardware-dependent calibration parameters: The parameters $a(k)$ and $b(k)$ were calibrated on the MN5 platform. Deploying the model on a different cluster therefore necessitates recalibration, although this only requires running the QuCSLO Calibrator (Section~\ref{subsec:impl-resources}) once on the target machine.

\noindent Beyond these limitations, we plan to \textbf{explore}:

1) Integration with persistent simulation services: Systems such as CUNQA~\cite{vazquezperez_cunqa_2025} deploy long-running services on fixed node allocations and support runtime backend switching, which avoids reprovisioning the service when the backend changes. Our framework is complementary, as it resizes the allocation to the requirements of each circuit. Combining both approaches could enable dynamic scaling of virtual quantum processing units (vQPUs) according to the active workload or real-time utilization metrics.

2) Support for quantum resource managers routing to diverse backends: Modern resource managers may dynamically route circuits, or the subcircuits produced by circuit cutting, to either classical simulators or physical QPUs~\cite{rocco_2025,garciaraigada2026,cipollini2026,10.1145/3731599.3767547}. Once a classical simulator is selected, the manager must still configure an efficient node count and qubit layout. Our placement and node-count models automate these decisions, eliminating the need for manual resource provisioning and circuit-specific preliminary sweeps.


\section{Conclusion}\label{sec:conclusion}
This work addresses two bottlenecks in distributed state-vector simulation: communication overhead caused by inefficient qubit placement, and rigid resource allocation. We introduce a placement optimizer that models the circuit as a weighted graph and uses simulated annealing to reduce the communication cut, which exactly counts the communicating gates, together with a node-count efficiency model that selects, before execution, the largest node count whose predicted degradation stays below a user-defined threshold. To the best of our knowledge, this is the first time the node count of a distributed simulation is derived from the circuit itself rather than fixed arbitrarily, set to the maximum allocation available, or found through a preliminary sweep. Experiments on synthetic and QASMBench circuits show speedups of up to 2.3$\times$ for circuits with sparse, structured interactions, and the model selects the same node count as the measurements in 83--92\% of cases. Implemented on QuEST and DMR, the automatic configuration matches the runtime of a greedy 64-node allocation for \emph{bv\_n30} at 1/36 of its node-hour cost, and reduces the cost of \emph{vqe\_uccsd\_n28} by 98\% without a preliminary sweep. Future work will integrate the framework with persistent simulation services and quantum resource managers.

\begin{acks}

The authors thank the Quantum Applications Group at EPCC, and in particular James Richings and Oliver Thomson Brown, for providing the \emph{QFT-epcc} circuit used in this work.

BSC researchers have been financially supported by: ``Barcelona Zettascale Laboratory (BZL)'' and ``QUANTUM ENIA project call - Quantum Spain project'' backed by the Ministry of Economic Affairs and Digital Transformation of the Spanish Government and by the European Union through the Recovery, Transformation and Resilience Plan - NextGenerationEU within the framework of the Digital Spain 2026 Agenda.
Antonio J. Peña was partially supported by the Ramón y Cajal fellowship RYC2020-030054-I funded by MCIN\slash AEI\slash 10.13039\slash 501100011033 and by ``ESF Investing in your future''.

The views and opinions expressed are solely those of the authors and do not necessarily reflect those of the European Union. Neither the European Union nor the European Commission can be held responsible for them.

Language polishing was performed using an AI language model, with subsequent thorough review and editing by the authors. The authors take full responsibility for the final manuscript content.
\end{acks}

\bibliographystyle{ACM-Reference-Format}
\bibliography{sample-base}

\end{document}